# Possibility of single biomolecular imaging with coherent amplification of weak scattering X-ray photons

Tsumoru Shintake

RIKEN SPring-8 Center, Harima Institute

1-1-1 Kouto, Sayo, Hyogo 679-5148, Japan

The number of photons available by coherent X-ray scattering from a single biomolecule is considerably less because of the extremely small elastic-scattering cross-section and low damage threshold. Even with a high X-ray flux of $3 \times 10^{12}$ photons per 100-nm-diameter spot and an ultrashort pulse of 10 fs driven by a future X-ray free electron laser (X-ray FEL), it has been predicted that only a few 100 photons will be available by scattering from a single lysozyme molecule. In observations of scattered X-rays on a detector, the transfer of energy from wave to matter is accompanied by the quantization of the photon energy. Unfortunately, X-rays have a considerably high photon energy of 12 keV at wavelengths of 1 Å, which is required for atomic resolution imaging. Therefore, the number of photoionization events is considerably less, which limits the resolution of the imaging of a single biomolecule. In this paper I propose a new method: instead of directly observing the photons scattered from the sample, we amplify the scattered waves by superimposing an intense coherent reference pump wave on it and record the resulting interference pattern using a planar X-ray detector, similar to technique followed in holography. Using a nanosized gold particle as a reference pump wave source, we can collect $10^4$~$10^5$ photons in single shot imaging where the signal from a single biomolecule is amplified and recorded as 2D diffraction intensity data. To recover the phase information and reconstruct the image, the iterative phase retrieval technique will be applicable. However, the technical difficulty is how to precisely reconstruct faint image of the single bio-molecular in Angstrom resolution, whose intensity is much lower than the bright gold particle linked to it. In order to solve this problem, I propose a new scheme that combines the iterative phase-retrieval on reference pump wave and digital Fourier transform holography.

## I. INTRODUCTION

Protein crystallography has advanced through the use of high-quality X-ray beams generated by synchrotron light sources [1], and especially with the development of higher-brightness X-ray beams produced by an undulator device placed in low-emittance electron storage rings [2]. The increase in the brightness of X-ray beams has made it possible to increase the flux incident on small protein crystals. Currently, crystals with a diameter of 100-μm are being routinely analyzed, and the analysis of 10-μm-diameter crystals is under investigation. It is comparatively easier to form high-quality small crystals than larger crystals with lesser number of defects.

However, crystallization of membrane proteins is extremely challenging because they require detergents to maintain them in isolation or solution; unfortunately, such detergents often interfere with the crystallization process. Membrane proteins form a large component of the genomes and include several proteins of great physiological importance, such as ion channels and receptors [3].

Recently, the practical applications of single-pass free-electron lasers (FELs) [4,5] at short wavelengths have increased as a result of long-term research and development of important devices and technologies such as low-emittance electron sources, accurate undulator devices and high-precision electron-beam control techniques. Currently, three large X-ray FELs emitting 1 Å X-ray laser beam are under construction and will be ready in the next few years [6–8].

Since X-ray FELs produce extremely high peak power with short pulse duration, the potential use of these devices to determine the structure of a single biomolecule or individual cells from scattered photons in one shot has been studied [9]. Through a computer simulation, which considers in detail the damage associated with photoabsorption, Auger electron emission, and their various effects, it has been demonstrated that we can collect coherently scattered photons from a single biomolecule using femtosecond pulses before the atoms start drifting due to the Coulomb forces. However, the number of available photons is quite limited. In a previous report, a maximum 682 photons have been predicted to be scattered from a single lysozyme molecule. In the calculation, however, unrealistic parameters have been assumed: $5 \times 10^{13}$ photons per pulse in a 100-nm-diameter spot with a 1-fs pulse duration at a wavelength of 1 Å. If we refer to the design parameters of the X-ray FELs currently under construction, the feasible parameters appear to be $3 \times 10^{12}$ photons or less per pulse in a 100-nm-diameter spot with a 10-fs pulse duration; with these parameters, only 100 photons would be scattered. This is an extremely small number, making it impossible to carry out single-shot imaging.

In this paper, initially, I discuss the double slit thought experiment in order to show the signal amplification phenomena associated with the interference effect. On the basis of the discussion, I propose a new method: the scattered photons are amplified by superimposing a reference pump wave and the resulting interference fringes are recorded on a planar X-ray detector, similar to the technique followed in holography. Using four slit thought experiment, I discuss the signal contents in the interference fringes. To recover image, I discuss the interactive phase retrieval technique and also a new scheme that combines the iterative phase-retrieval on low-pass filtered data and digital Fourier transform holography.

## II. DOUBLE-SLIT THOUGHT EXPERIMENT

Figure 1 shows a modified version of Young's double-slit experiment. As compared slit $S_1$, the width of slit $S_2$ is made considerably small so that the intensity (and photon flux) of light passing through $S_2$ is considerably lower than that passing through $S_1$. The wave functions of the two beams arriving at the screen are expressed as follows:

$$\psi_1 = A_1 \cdot \exp i(\omega t - kz + \phi_1), \qquad (1)$$

$$\psi_2 = A_2 \cdot \exp i(\omega t - kz + \phi_2). \qquad (2)$$

These amplitudes are normalized such that the conjugate product $\psi^*\psi = |\psi|^2$ is equal to the photon flux of each beam. Therefore, the probability of finding a photon in the beams is proportional to $I_1 = \psi_1^*\psi_1$ for beam-1 and $I_2 = \psi_2^*\psi_2$ for beam-2.

The two waves interfere and the resulting interference fringes, which are observed on the screen, can be represented by the linear summation of the two wave functions, $\psi = \psi_1 + \psi_2$. The probability of finding a photon on the screen is given by

$$\begin{aligned} I = |\psi|^2 &= \psi^*\psi = \left(\psi_1^* + \psi_2^*\right) \cdot \left(\psi_1 + \psi_2\right) \\ &= |\psi_1|^2 + 2|\psi_1||\psi_2|\cos(\phi_1 - \phi_2) + |\psi_2|^2 \\ &= I_1 + 2\sqrt{I_1 I_2}\cos(\phi_1 - \phi_2) + I_2 \end{aligned} \qquad (3)$$

In the X-ray diffraction experiment, $\psi_1$ represents the reference pump wave and $\psi_2$ represents the diffracted wave from the single biomolecule. The second term in Eq. (3)

represents the interference fringes, which contain information about the second slit, i.e., its width, intensity of the light beam passing through it, or its distance from the first slit. If there are multiple slits, there will be multiple frequency components in Eq. (3), which can be analyzed by Fourier transformation.

Next, we discuss information that is retrievable from fringe contrast. From the maximum and minimum amplitudes given in Eq. (3), we determine the modulation depth as follows:

$$M = \frac{I_{max} - I_{min}}{I_{max} + I_{min}} = \frac{2\sqrt{I_2/I_1}}{1 + I_2/I_1}. \qquad (4)$$

Figure 2 shows the modulation depth (or visibility) as a function of the intensity ratio $I_2/I_1$. The modulation depth has a maximum value of 1 when the intensities are equal, i.e., $I_2 = I_1$. By decreasing $I_2$, the modulation depth decreases, although the rate of this decrease is slower than a first-order dependence, which indicates that the modulation depth or fringe contrast is visible even for very small values of $I_2$.

This phenomenon has been widely used to detect weak signal in the radio frequency communication system and also in optical laser system; it is called heterodyne detection method, or homodyne detection method when frequencies of the two beams are identical. The intense beam-1 is a local oscillator in the case of the radio frequency applications, or a reference pump beam in the optical laser system. In the X-ray crystallography, there is a very important technique using this phenomena; it is called the isomorphous replacement method [10]. One or a few heavy atoms attached to a large protein molecule provide sufficient change intensities of the diffractions, from which phase of Bragg's diffraction spots can be determined. The effect of intensity change was estimated by Crick and Magdoff [11], and they found the root mean square relation, which is basically same as eq. (4).

In this paper, I propose to place a nanosized heavy atom nearby a single bio-molecule molecule, by which the diffraction will be amplified and also its phase can be determined. If we count number of the photons directly, we will observe only a few 100 of photons scattered from the single biomolecular. We need a least ten times or more photons for the biomolecular imaging. The basic concept of quantum mechanics of wave–particle duality states that all matter exhibits both wave- and particle-like behavior [12,13]. We can choose not to directly observe the X-ray photons scattered from the single biomolecule, and allow them to propagate in free space as a wave and overlap with coherent and more intense waves from $S_1$. Since the resultant wave has considerably large amplitude, a large number of photoionization events occur in the detector; thus, we can obtain more precise information about the biomolecule, which is recorded in the amplitude modulation of the interference fringes.

### III. PROPOSED SETUP OF HOLOGRAPHIC RECORDING

As shown in Fig. 3, we place a single biomolecule along with a small nanosized gold particle, which is used as the scattering source for the reference pump wave, in the path of an intense X-ray beam generated by an X-ray FEL. As discussed later, the size of the scattering source has to be small for better separation on frequency components in the X-ray scattering data, while we need high pump wave intensity, therefore we choose heavy atom with high-$Z$ material: since atomic scattering form-factor scales as $Z^2$. The gold particle and single biomolecule correspond to the wide and narrow slits, respectively, in the double-slit experiment described previously. The scattered waves from these two objects interfere and form an interference pattern on the planar detector. There are various components in the interference pattern; of these, the simplest pattern is a periodic sine wave pattern associated with the two-wave interference, whose width is given by

$$p = \frac{L}{D}\lambda \qquad (5)$$

where $L$ is distance between the objects and the detector; $D$, the distance between the single biomolecule and the gold particle; and $\lambda$, the X-ray wavelength. We assume the planar detector (100 mm × 100 mm) with 256 × 256 pixel is placed at a distance of 100 mm from the sample (the resolution at rim corresponds to 2 Å for a wavelength of 1 Å). The interference fringe width should be 1 mm or larger for the pattern to be clearly visible on the detector. From Eq. (5), we find that the distance between the two objects must be 10 nm or less, which is less than the coherence length of the X-ray beam emitted from the FEL.

Here, we consider as an example a single lysozyme molecule, which has a mass of 14.4 kDa. It is one of the standard enzymes, with a size of approximately 45 × 30 × 30 Å. It consists mainly of carbon, oxygen, and nitrogen atoms. In order to observe this structure at the spatial resolution around 2 Å, the number of Fourier components required to represent the 2D-projection image of this structure becomes roughly 50 × 50, where a factor of 2~3 for the over sampling factor is assumed. There must be at least ten photon events for each Fourier component; thus the required total number of photons becomes $50 \times 50 \times 10 = 2.5 \times 10^4$. This is fairly rough estimate for this small sample case. When the size of the sample becomes larger and more complex, larger number of pixels and X-ray photons are required.

However, a previous study [9] has predicted the number of available photons to be around only a few 100 from scattering by a single lysozyme molecule, which is considerably less than the number of photons required for imaging. We need 100~1000 times more flux in one shot.

Now, we roughly estimate the required number of atoms in the gold particle. The elastic scattering cross section is nearly proportional to the square of the atomic number $Z$. The scattering power decreases with the scattering angle and decreases rapidly for a low-$Z$ material [14]; we neglect this effect for simplicity. The number of scattering photons under the same flux becomes

$$\frac{n_{p.Au}}{n_{p.C}} = \frac{N_{Au}\sigma_{Au}}{N_C\sigma_C} = \frac{N_{Au}Z_{Au}^2}{N_C Z_C^2}. \qquad (6)$$

We approximate the lysozyme molecule as a 1000-atom carbon cluster, denoted by the subscript C. Using $Z_C = 6$, $N_C = 1000$, $Z_{Au} = 79$, and the same number of Au atoms as the lysozyme, i.e., $N_{Au} = 1000$, we find that intensity ratio: $n_{p.Au}/n_{p.C} \approx 200$. A gold particle in the metallic phase with a diameter of less than 3 nm contains 1000 atoms. Using latest advanced gold-labeling technology [15], we can bind a nanosized gold particle to various types of biomolecules such as proteins, lipids, or ATP. Products such as Nanogold are manufactured using a well-established technology and are commercially available in diameters from 1.4 to 40 nm. The linking arm on the gold particle can be made specifically to react with thiols; thus, the location of the link to the biomolecule is well defined.

When an intense X-rays beam with a flux of $3 \times 10^{12}$ photons per pulse in a 100-nm-diameter spot is incident on the gold particle linked to a single lysozyme molecule, as shown in Fig. 4, we obtain $2 \times 10^4$ photons on the X-ray detector. To provide uniform illumination of reference pump wave within the planar X-ray detector, a non-crystal structure will be suitable for the gold particle. R&D will be required to optimize its metallic structure (amorphous, quasi-crystal gold, glassy alloy, etc.), but for simplicity, here we assume a perfect gold particle with random atom distribution.

## IV. FOUR-SLIT THOUGHT EXPERIMENT

The experimental setup shown in Fig. 3 is identical to that used for Fourier transform holography or lens less holographic imaging [16, 17]; only the reference wave used in these experiments is different.

The Fourier transform holography uses spherical waves as reference waves to record the phase of an object wave. The intensity of the reference wave is chosen to be comparable to that of the object wave to obtain the best contrast. For better image quality, the size of the reference wave source has to be considerably smaller than the object. The image-recovery process in Fourier transform holography is simple; basically, an inverse Fourier transform performed on the obtained intensity data provides the real image. The technique is well established, and Fourier transform holography has been successfully applied to the observation of magnetic nanostructures in thin films using soft X-rays [18].

In our new method, the intensity of the reference wave is considerably higher; as a result, the reference wave amplifies the weak signal and also provides a phase reference. However, the diameter of the gold particle is comparable to that of the single biomolecule and it is not a point source. Hence, the scattered X-rays cannot be simple spherical waves.

In fact, they form a speckle pattern, whose phase and amplitude are not known. Therefore, simple holographic image processing cannot be applied in this case.

To solve this problem, I devised a thought experiment with each slit in the double-slit experiment having internal structures, i.e., there are in total four slits as shown at bottom of Fig. 4. The gold particle, internal structure or an imperfection of the gold particle, the sample, and the internal structure of the sample function as the first, second, third, and fourth slits, respectively. This is the simplified model of a gold-labeled molecule. When coherent light is incident, the probability of finding a photon on the screen is given by

$$\begin{aligned} I = |\psi|^2 = \psi^*\psi = \sum_{i=1}^{4} \psi_i^* \cdot \sum_{j=1}^{4} \psi_j \\ = I_1 + 2\sqrt{I_1 I_2} \cos\Delta\phi_{12} + 2\sqrt{I_1 I_3} \cos\Delta\phi_{13} + 2\sqrt{I_1 I_4} \cos\Delta\phi_{14} \quad (7) \\ + I_2 + 2\sqrt{I_2 I_3} \cos\Delta\phi_{23} + 2\sqrt{I_2 I_4} \cos\Delta\phi_{24} \\ + I_3 + 2\sqrt{I_3 I_4} \cos\Delta\phi_{34} + I_4 \end{aligned}$$

where $\Delta\phi_{ij} = \phi_j - \phi_i$ is the phase difference given by

$$\Delta\phi_{ij} = \phi_j - \phi_i = \frac{2\pi d_{ij}}{\lambda} \cdot \sin(2\theta) \qquad (8)$$

Here, $I_i$ represents the flux from the i-th slit, $2\theta$ is the scattering angle (as defined in crystallography), and $d_{ij}$ is the distance between the $i$- and $j$-th slits. The slit locations and flux ratio suitable for our example case shown in Fig. 4 are as follows: $d_{12}$ = 1.5 nm, $d_{23}$ = 8.5 nm, $d_{34}$ = 2 nm, $I_1$ = 200, $I_2$ = 10, $I_3$ = 1, $I_4$ = 0.1, and $\lambda$ = 1 Å. Figure 5 (a) shows the flux density distribution estimated by Eq. (7). The distribution is considerably complicated because the interference fringes are formed by the interference of four waves.

In Fig. 5(c), the curve at the bottom indicates the scattered wave from the biomolecule when it is directly observed without using the reference pump wave. It is a very weak signal with a relative intensity of approximately 1. In practice, the signal is quantized by the photon energy, leading to the loss of detailed information. The dashed curve (magnified by 10) also shows an interference pattern, which represents the internal structure of biomolecule; our aim is to study this pattern. By superposition of the reference pump wave, the signal wave is amplified, and the interference pattern obtained is recorded. In order to demonstrate the amplification effect clearly, the reference pump wave is assumed to be perfect with $\psi_2 = 0$. In Fig. 5(c), the curve at top shows the amplified signal, which is recorded by the amplitude modulation of the interference pattern. From Eq. (7), we find that

$$I = I_1 + 2\sqrt{I_1 I_3} \cos \Delta \phi_{13} + 2\sqrt{I_1 I_4} \cos \Delta \phi_{14}$$
$$+ I_3 + 2\sqrt{I_3 I_4} \cos \Delta \phi_{34} + I_4 \quad (9)$$

In our experiment, the reference wave has considerably higher intensity than others; $I_1 \gg I_3 > I_4$. Hence, we can eliminate the last three terms (direct wave) and Eq. (9) becomes

$$I \approx I_1 + 2\sqrt{I_1 I_3} \cos \Delta \phi_{13} + 2\sqrt{I_1 I_4} \cos \Delta \phi_{14} \quad (10)$$

These two amplified signals have different frequencies, which results in the formation of a beat wave at $\Delta \phi_{14} - \Delta \phi_{13} = \Delta \phi_{34}$, which has a frequency equal to the interference signal of the direct sample signal. As discussed later the sample signal is amplified and recorded as the beat wave. The signal gain is given by

$$G = \frac{\text{modulation in amplified signal}}{\text{modulation in direct signal}}$$

$$= \frac{2\left(\sqrt{I_1 I_3} + \sqrt{I_1 I_4}\right) - 2\left(\sqrt{I_1 I_3} - \sqrt{I_1 I_4}\right)}{4\sqrt{I_3 I_4}} \quad (11)$$

$$= \frac{\sqrt{I_1 I_4}}{\sqrt{I_3 I_4}} = \sqrt{I_1 / I_3}$$

In this example, $G = \sqrt{200} = 14.1$.

The Bragg diffraction is based on the same gain phenomenon. In the case of a crystal, small components are coherently scattered by several atoms interacting constructively and the wave is "amplified" and finally emitted in a certain direction that satisfies the Bragg condition. There is no net energy gain in this process. Constructive interference takes place depending on the phase-matching condition and allows reflection into the singular Bragg spots, which are significantly brighter than the incoherent background.

The signal to noise ratio $S/N$ is significantly improved as follows. We assume the noise is caused by statistical fluctuations associated with the quantization process at photoionization. The improvement in $S/N$ can be estimated by

$$\frac{S/N|_{\text{holography}}}{S/N|_{\text{direct}}} = \frac{\sqrt{I_1 I_4}/\left(\sqrt{I_1}/I_1\right)}{\sqrt{I_3 I_4}/\left(\sqrt{I_3}/I_3\right)} = \frac{I_1}{I_3} = 200. \quad (12)$$

Therefore, a more accurate image can be obtained with this method. It should be noted that we do not detect the photons directly after it is scattered from the biomolecule; we detect it by its interference with the superimposed reference pump wave, which yields an amplified signal, resulting in considerably high power and lower statistical deviations associated with the quantization process.

Figure 5(b) shows the reference pump wave, which has interference term of $\psi_1 + \psi_2$. In practice, it forms a speckle pattern due to the random distribution of the gold atoms. Problem is that the amplified signal is influenced by the randomness in the speckle. To employ Fourier transform holography, we have to determine the phase and amplitude of the reference pump wave.

In Eq. (7), under the condition $I_1 > I_2 \gg I_3 > I_4$, we can eliminate the direct waves from the sample as follows:

$$I = I_1 + 2\sqrt{I_1 I_2} \cos \Delta\phi_{12} + 2\sqrt{I_1 I_3} \cos \Delta\phi_{13} + 2\sqrt{I_1 I_4} \cos \Delta\phi_{14} \\ + I_2 + 2\sqrt{I_2 I_3} \cos \Delta\phi_{23} + 2\sqrt{I_2 I_4} \cos \Delta\phi_{24} \tag{13}$$

In this equation, the information from the sample is contained in the interference patterns as $\psi_1\psi_3$, $\psi_1\psi_4$, $\psi_2\psi_3$, and $\psi_2\psi_4$.

In order to measure the spatial frequency components of the interference pattern, we define the angular frequency as follows:

$$\phi'_{ij} = \frac{\partial \Delta\phi_{ij}}{\partial 2\theta} \approx \frac{2\pi d_{ij}}{\lambda} \tag{14}$$

The angular frequency is proportional to the distance $d_{ij}$ or the diameter of the particle. If we multiply the angular frequency with the distance from the object to the planar detector, we obtain the spatial frequency. The maximum angular frequency of a speckle from the gold particle is determined by interference of two atoms at left-edge and right-edge, that is, $\phi'_{sp.max} \approx 2\pi \times 2d_{12}/\lambda = 2\pi D_{Au}/\lambda$, where $D_{Au}$ is the gold particle diameter. If the gap between the linked structure and the gold particle is designed to be larger than the gold diameter, i.e., $d_{23} > D_{Au}$, the angular frequency of the interference fringes of the signal wave exceeds the speckle angular frequency, i.e., $\phi'_{13}, \phi'_{14}, \phi'_{23}, \phi'_{24} \geq \phi'_{sp.max}$. Therefore, if we use a low-pass spatial filter, we may be able to separate the speckle pattern from the high frequency components.

# V. ITERATIVE PHSE RETRIEVAL AND IMAGE RECONSTRUCTION

In the coherent X-ray diffraction imaging, or lens-less X-ray diffraction microscopy, the X-ray diffractions from all electrons in the sample create interference pattern and recorded as 2D diffraction intensity data, which usually takes a fairly complicated and random image like a speckle pattern. Recently, the digital signal processing technology has made it possible to reconstruct the object image from such diffraction data. The technique is called the iterative phase retrieval or the over sampling method [19–23].

As seen in the previous section, signal from the single bio-molecule is amplified on the intense reference pump wave and recorded as an interference pattern. By applying iterative phase retrieval technique, we may recover the phase of the diffraction data which includes the interference pattern between two objects (gold particle and the single molecule), and we may reconstruct the images of the single molecular linked with the gold particle as one object.

Benefit and drawback of this method are discussed below.

(1) Benefit of using intense reference pump wave is that the S/N ratio has been much improved as shown in eq. (12), and resulting in improving image quality. We may use even higher reference pump wave, until it hits the limit of the dynamic range in the X-ray detector.

(2) The ultimate resolution of the diffraction imaging is defined by the maximum diffraction angle at the edge of the planer detector: $d \approx \lambda / 2 \sin \theta_{max}$. For example, using X-ray of 1 Å wavelength and the maximum diffraction angle: $2\theta_{max} = 30°$, the resolution limit becomes 2 Å. However, in case of the single molecular imaging, the available number of photons is quite limited; as a result the practical resolution becomes much worse due to the background noise and large statistical fluctuations on number of photons in each pixel. By introducing intense pump wave, the S/N ratio can be much improved as shown in eq. (12), therefore the practical resolution will be improved.

(3) By introducing the gold particle, the total size of the sample becomes larger, which makes the maximum spatial frequency in the diffraction data much higher. To correctly capture this image, we have to use finer pixel size in the planer detector, which reflects back to lower number of photons in each pixel. In the example case of Fig. 4, the full size of the sample is about 4 times larger than the single lysozyme, therefore we have to use 4 x 4 = 16 times more pixels, and the number of photons in each pixel becomes 16 times lower. This is one drawback of this method, but it will be compensated by the much higher gain of S/N ratio; i.e., 200 times in the present example.

(4) In case of larger molecule such as the single protein molecule or single viral capsid, the diffraction pattern becomes much more complicated and extends to higher spatial frequency, which potentially makes the iterative phase retrieval process unstable and unreliable.

(5) Even if using most advanced phase retrieval algorism, always a few % of discrepancies are found in the retrieved images [24]. Therefore, it will be difficult to study on detail structure of the single bio-molecular in the retrieved image, whose intensity is at least 100 times lower than the bright gold particle.

In order to solve this problem, I propose a new scheme that combines the iterative phase-retrieval on low-pass filtered data and digital Fourier transform holography in the following sections.

## VI. HOLOGRAPHIC IMAGE RECONSTRUCTION

Here we consider two groups of atoms, group I and group II, as shown in Fig. 6. Group I has a higher scattering power than group II. The resultant diffracted wave is given as follows:

$$\begin{aligned}
\psi &= \sum_{k=1}^{N_1} \psi_{1k} + \sum_{l=1}^{N_2} \psi_{2l} \\
&= \sum_{k=1}^{N_1} f_k \exp[2\pi i \mathbf{r}_{1k} \cdot \mathbf{S}] + \sum_{l=1}^{N_2} f_l \exp[2\pi i (\mathbf{R}_{12} + \mathbf{r}_{2l}) \cdot \mathbf{S}] \\
&= \sum_{k=1}^{N_1} f_k \exp[2\pi i \mathbf{r}_{1k} \cdot \mathbf{S}] + \exp[2\pi i \mathbf{R}_{12} \cdot \mathbf{S}] \cdot \sum_{l=1}^{N_2} f_l \exp[2\pi i \mathbf{r}_{2l} \cdot \mathbf{S}] \\
&= R + O
\end{aligned} \quad (15)$$

where $f$ is the atomic scattering factor and $\mathbf{S}$ is the scattering vector given by $|\mathbf{S}| = 2\sin\theta/\lambda$. The origin in each group, $G_1$ and $G_2$ is defined as the center of charge density. $\mathbf{r}_{1k}$ and $\mathbf{r}_{2l}$ represent the particle distance from the origins, and $\mathbf{R}_{12}$ is the distance between the origins. $R$ and $O$ denote the reference and object waves; here, we follow the same notation used in holography. The intensity of the diffracted wave becomes

$$\begin{aligned}\psi^*\psi &= (R+O)^*(R+O) \\ &= R^*R + R^*O + O^*R + O^*O \quad (16)\\ &\approx R^*R + R^*O + O^*R\end{aligned}$$

We eliminate the last term because the contribution of the direct signal $O^*O$ is considerably smaller as compared to other signals. As given in Eq. (15), the diffracted wave from group II has the phase term $\exp[2\pi i \mathbf{R}_{12} \cdot \mathbf{S}]$, which has considerably high spatial frequency components.

Figure 7 shows the frequency components in the diffracted waves. If there is a gap between the reference wave and the object wave, we may separate those signals by means of frequency filters. The condition for this

$$g_{link} > D_{Au} \quad (17)$$

where $g_{link}$ is the gap between the right end of the gold particle and the left end of the sample. In our example shown in Fig. 4, the gap $g_{link}$ has to be longer than 3 nm.

If we use a band-pass filer denoted BPF in Fig. 7, we can obtain the interference term $R^*O + O^*R$. However, it contains unknown phase terms $R^*$ and $R$. To find phase, we apply the iterative phase-retrieval method.

Initially, we use a low-pass filter, denoted by a LPF in Fig. 7, and eliminate the high frequency components. If we chose turn over frequency of low-pass filter at the gap, we obtain the approximated reference wave as follows.

$$LPF\left[\psi^*\psi\right] \simeq \sum_{k=1}^{N_1}\psi_{1k}^* \sum_{k=1}^{N_1}\psi_{1k} = R^*R \quad (18)$$

Then we apply the iterative phase-retrieval method on $R^*R$ data. Since reference pump wave contains a large number of photons diffracted from the gold particle, and it is to some extent an isolated object and spatial frequency is much lower than the maximum sampling frequency, which is right end in Fig. 7, therefore the over sampling ratio becomes much higher, and the iterative process will always converge, and provide accurate solutions [24]. Error associated with digital filter is discussed in the next section.

Once we determine the reference wave, the image recovery process of the single biomolecule is simple. From Eq. (16),

$$\frac{\psi^*\psi - R^*R}{R^*} = O + RO^*/R^* \qquad (19)$$

where the second term is a conjugate image, which will create a ghost image opposite to the gold particle and does not deteriorate the real image [26]. The subtraction $\psi^*\psi - R^*R$ plays important roll in holographic image reconstruction, which erases the intense reference wave and the residual becomes only the interference term. Therefore the inverse Fourier transform of eq. (20) produces clean image of the single molecule. On the other hand, applying the iterative phase retrieval technique on both objects at once, even a small phase error will create image leakage from the bright gold particle, which will contaminate the single molecular image. The operator $1/R^*$ on the left-hand side of the equation applies the amplitude normalization and phase correction to the diffraction pattern for a non-uniform reference wave distribution (correction of the speckle pattern formed from the gold particle).

To reconstruct the electron density map in the sample, we apply the inverse Fourier transform as follows [25]:

$$\rho(\mathbf{r}) = \int (\psi^*\psi - R^*R)/R^* \cdot \exp[-2\pi i \mathbf{r} \cdot \mathbf{S}] \cdot d\mathbf{S} \qquad (20)$$

where $\rho(\mathbf{r})$ includes the atomic scattering factor $f$. The approximated 2D projection of the object is obtained as follows:

$$\rho(x,y) = \int_{S_y}\int_{S_x} (\psi^*\psi - R^*R)/R^* \cdot \exp[-2\pi i(x \cdot S_x + y \cdot S_y)] \cdot dS_x dS_y \quad (21)$$

We may use the band-pass filter, which is denoted as BPF in Fig. 7, to select the object wave and reject the background noise [27, 28] and parasitic diffractions coming from the upstream of X-ray beam line.

In actual experiments, a large number of biomolecules with gold labeling must be injected into a beam line formed by a spraying technique and illuminated by an X-ray FEL. The orientation of the biomolecules will be random. The experiment is carried out as follows (refer Fig. 8):

(1) Image selection: Applying a fast Fourier transform (FFT) on 2D data obtained from the planar detector, we obtain a peak corresponding to $\mathbf{R}_{12} \cdot \mathbf{S} = 1$, from which we determine the axis vector $\mathbf{R}_{12}$. We have to choose the optimum diffraction images from a large number of images of randomly oriented molecules. The best image will have the largest $\mathbf{R}_{12}$ vector, whose length must be same as that of the designed value of link structure of gold particle, and highest contrast; it will be achieved when the gold label and biomolecule lie in a plane normal to the beam axis. A few degrees of angle error can be corrected in step-(5) by tilting image numerically.

(2) Retrieval of reference pump wave: The low-pass filter (2D smoothing) is applied to the scattering X-ray data to eliminate the object waves, followed by iterative phase-retrieval process. Using the filtered scattering data as the amplitude and the retrieved phase, we compute the reference pump wave.

(3) Acquiring 2D projection image of the single molecule: A numerical operation is employed to extract the object wave by Eq. (19), and inverse FFT of Eq. (21) is applied to obtain the approximated 2D projected image.

(4) Rotation of axis: The axis of the retrieved images is aligned. To find the azimuthal rotation angle, an additional smaller gold label can also be used as an angle marker.

(5) 3D structure reconstruction: By combining electron densities give by Eq. (20) according to the azimuthally angle, we have 3D electron density map. Averaging over many numbers of data will improve resolution.

The hologram obtained by Eq. (19) contains depth information, i.e., it can represent 3D images just as common holograms. We may tilt single molecular image numerically [28], while maximum rotation angle is restricted within fraction of solid angle of planar detector acceptance. However, this feature will enable us to easily assemble 3D structures from a large number of 2D projection images by rotation and interpolation.

## VII. DISCUSSIONS

Once we resolve the reference wave from the diffraction data, we can apply the holographic imaging scheme which makes assembling process of the 3D-structure much easier as discussed in the previous section. Here I discuss on error associated with the digital filter used to obtain the reference wave. The low pass filter (LPF) is not perfect, thus some signal of $R^*O + O^*R$ will leak out and mix with $R^*R$, or LPF cut some amount of the high frequency part of $R^*R$. We may chose the cutoff frequency of LPF much closer to the interference signal $R^*O + O^*R$ or within it, thus we do not loose high frequency edge of the reference wave, which contain geometrical information of non-spherical surface structure of the gold particle. In the iterative phase retrieval technique, it is known that iteration becomes unstable for a symmetric structure. Therefore it is better to use non-spherical structure, and also the edge should not be rounded by LPF. In Fig. 4, the gold particle is drawn as a spherical ball, but in practice the nanosized gold particles take various non-spherical geometries.

On the other hand, the leakage signal of $R^*O + O^*R$ becomes a part of the reference wave, which means the gold particle having a tail structure of the link and the edge part of the single molecular. All of them become the reference wave source. We have been discussing wave interference between two groups of atoms. However in the experiment wavelets from all atoms interfere each others at once. Thus we may rewrite eq. (15) as follows.

$$\begin{aligned} \psi &= \sum_{n=1}^{N_1+N_2} \psi_n \\ &= R + O \\ &= (R+\delta) + (O-\delta) \end{aligned} \quad (22)$$

where $\delta$ represents the error component in the filtering, in practice the link structure of the gold particle and edge of the single molecule. After processing the numerical operation of eq. (19), it subtracts more by $\delta$, thus the edge part of the single molecular will be missing from the reconstructed image, which can be found as a part of the gold particle.

The tail on the gold particle will contribute to change the phase of the reference wave, but it does not introduce geometrical error on the reconstructed image. Because the phase distribution of each group will be differing according to the way of grouping, but the diffraction pattern from all particles does not change as long as eq. (22) being kept. The operation of eq. (19) always satisfies this constraint.

We have to note that if the gold particle has crystal structure, the reference wave will be concentrated onto the Bragg's diffraction spots, and at other locations the intensity becomes almost zero, thus the numerical operation of $1/R^*$ on left side of eq. (19) becomes unstable and introduces substantial error. Therefore, we need to find optimum structure of the gold particle, i.e., random atom distribution to provide uniform illumination, while it will be random speckle pattern. Intensive R&D work will be required to prepare such a gold particle, and also to test numerical simulations which carefully take into account those details, and optimize the numerical procedures.

## VIII. CONCLUSIONS

In this paper, I have discussed the potential of a new scheme in determining the atomic structure of a single biomolecule using an intense coherent X-ray beam from an X-ray FEL. A gold particle linked to the biomolecule functions as a scattering source for a reference pump wave, which amplifies the weak diffracted wave, and a large flux X-ray photons is incident on the detector; the scattered photons carry sufficient structural information that can be recorded by a holography technique. To reconstruct an image from the X-ray diffraction data, I have devised a scheme that combines the iterative phase-retrieval technique and digital Fourier transform holography. This scheme can be used to reconstruct the image of a single biomolecule. Using a combination of advanced gold labeling technology and intense X-ray pulses from X-ray FELs, we will be able to study the structure of various proteins with a resolution close to few angstroms by this method, without crystallization. Detailed research and development will be required to optimize the gold particle structure (amorphous, quasi-crystaline, glassy alloy, etc.), and also the radiation damage on the gold particle. A proof of principle experiment will be required when the X-ray FELs emitting 1 Å X-ray laser beam are ready.

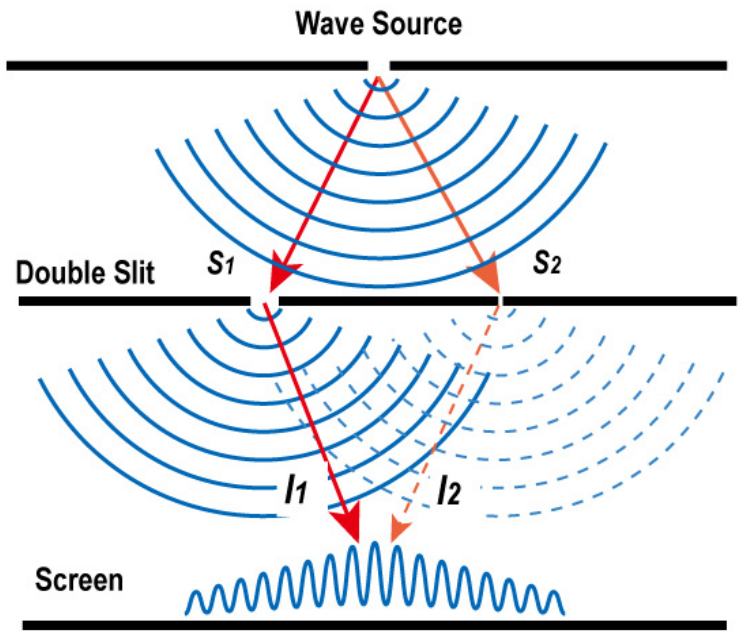

FIG. 1 (color online). Modified double-slit experiment

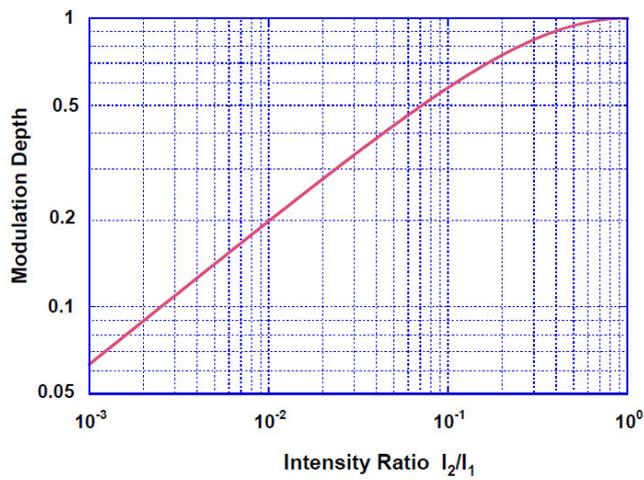

FIG. 2 (color online). Modulation depth of interference fringe from the modified double slit experiment.

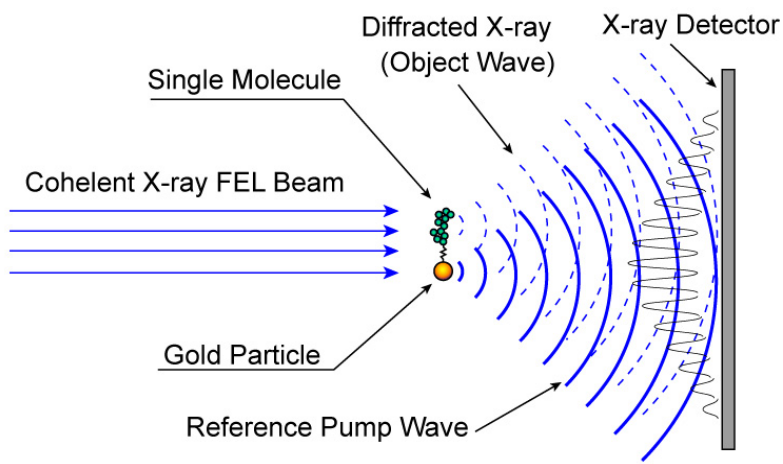

FIG. 3 (color online). Holographic recording for single-molecule imaging.

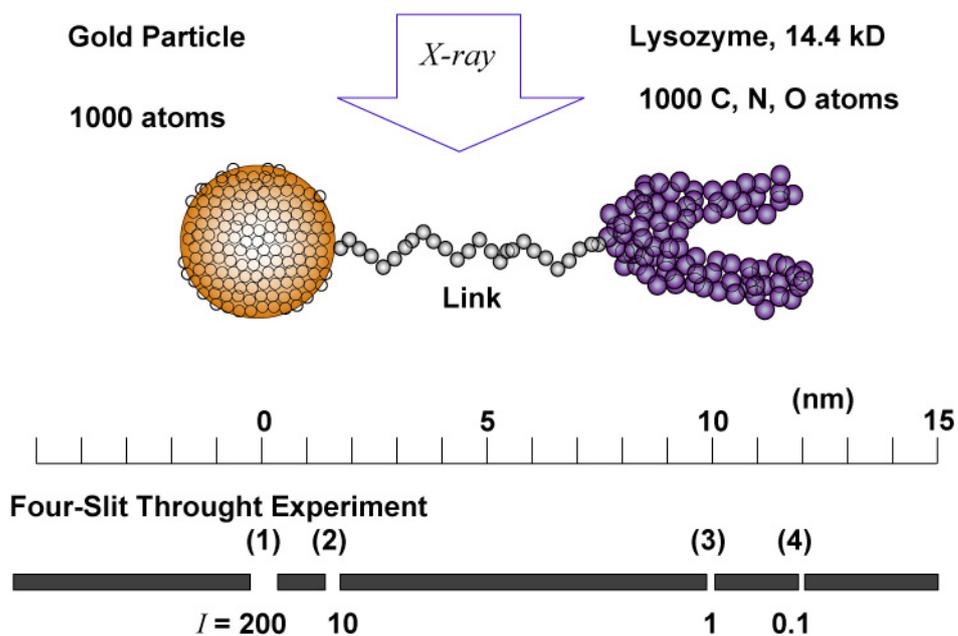

FIG. 4 (color online). Conceptual drawing of single lysozyme molecule linked to gold particle (diameter of atoms is not drawn to scale). The gold particle creates 200 times more coherent X-ray scattering than single lysozyme molecule. The bar at the bottom represents four-slit thought experiment.

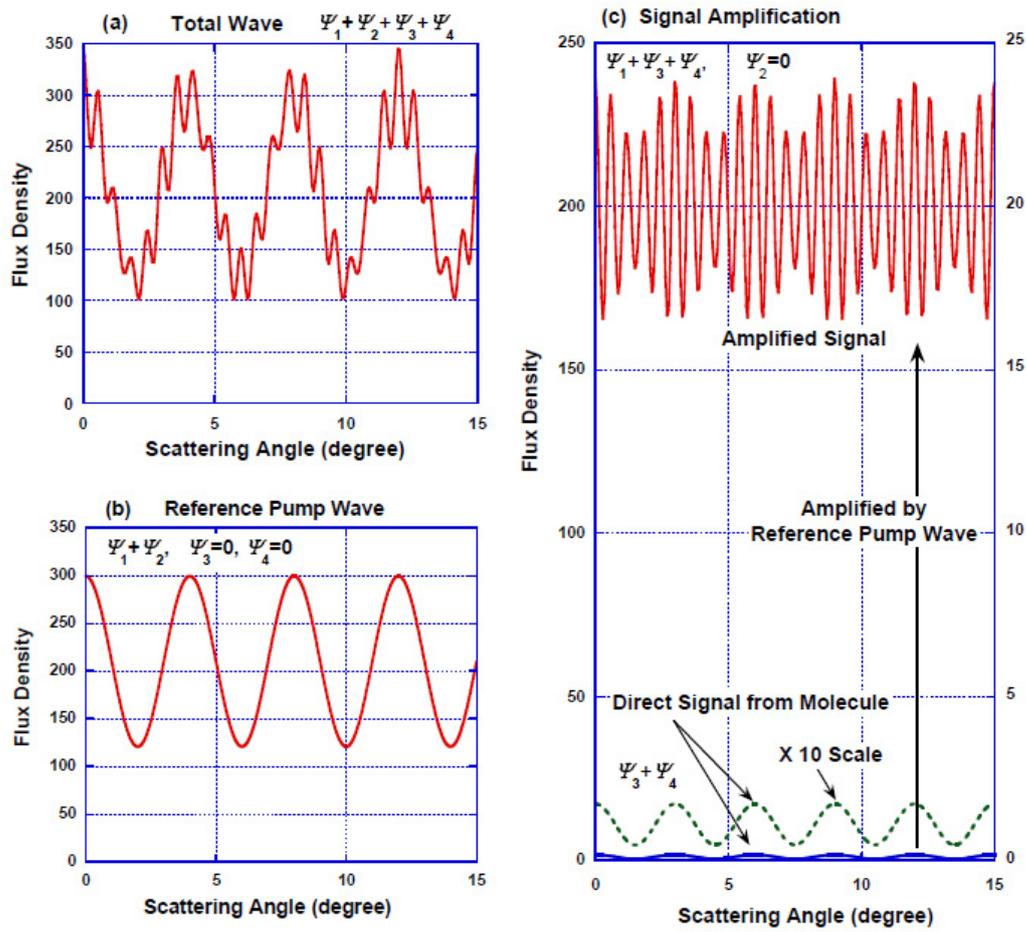

FIG. 5 (color online). Four slits thought experiment.

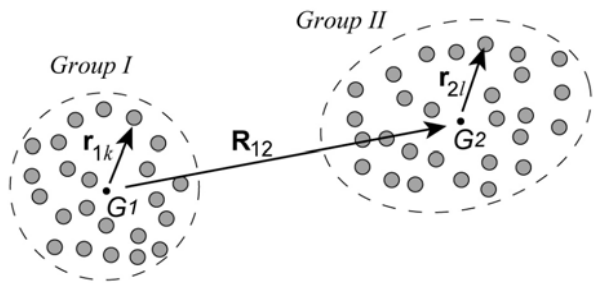

FIG. 6. Two group model of atoms. $G_1$ and $G_2$ are the mean centers of electron charge in each group

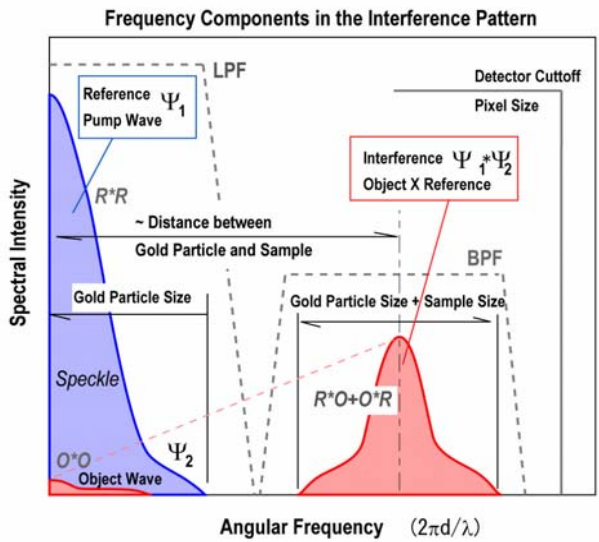

FIG. 7 (color online). Angular frequency components in the interference pattern: $\phi' = 2\pi d / \lambda$.

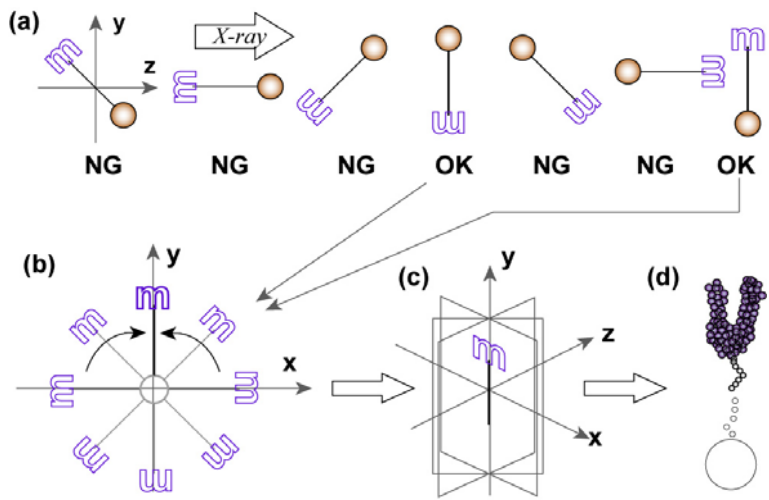

FIG. 8 (color online). Creating 3D structure from randomly oriented molecules. (a) Choose diffraction images which the gold particle and biomolecule lie in the xy-plane, (b) rotate image on xy-plane to align on y-axis, (c) assemble many images according to azimuthally angle, (d) created 3D structure.